\documentclass[pdflatex,sn-mathphys]{sn-jnl}

\jyear{2022}%

\theoremstyle{thmstyleone}%
\theoremstyle{thmstyletwo}%

\theoremstyle{thmstylethree}%

\begin{document}

\title[Article Title]{A Digital Alkali Spin Maser}


\author*[1]{\fnm{Stuart} \sur{Ingleby}}\email{stuart.ingleby@strath.ac.uk}

\author[1]{\fnm{Paul} \sur{Griffin}}\email{paul.griffin@strath.ac.uk}

\author[1]{\fnm{Terry} \sur{Dyer}}\email{terry.dyer@strath.ac.uk}


\author[1]{\fnm{Marcin} \sur{Mrozowski}}\email{marcin.mrozowski@strath.ac.uk}

\author[1]{\fnm{Erling} \sur{Riis}}\email{e.riis@strath.ac.uk}

\affil*[1]{\orgdiv{Department of Physics}, \orgname{SUPA, Strathclyde University}, \orgaddress{\street{107 Rottenrow East}, \city{Glasgow}, \postcode{G4 0NG}, \state{Lanarkshire}, \country{UK}}}

\abstract{Self-oscillating atomic magnetometers, in which the precession of atomic spins in a magnetic field is driven by resonant modulation, offer high sensitivity and dynamic range. Phase-coherent feedback from the detected signal to the applied modulation creates a resonant spin maser system, highly responsive to changes in the background magnetic field. Here we show a system in which the phase condition for resonant precession is met by digital signal processing integrated into the maser feedback loop. This system uses a modest chip-scale laser and mass-produced dual-pass caesium vapour cell and operates in a 50~$\mu$T field, making it a suitable technology for portable measurements of the geophysical magnetic field. We demonstrate a Cram\'er-Rao lower bound-limited resolution of 50~fT at 1~s sampling cadence, and a sensor bandwidth of 10~kHz. This device also represents an important class of atomic system in which low-latency digital processing forms an integral part of a coherently-driven quantum system. }




\maketitle

\section{Introduction}\label{sec_intro}

The generation of radio-frequency modulation in the polarisation or intensity of light transmitted through a precessing ensemble of optically pumped atomic spins is a phenomenon exploited in optical magnetometry \cite{Bloom1962,Lucivero2021}. The measured spin precession results from Zeeman splitting of the atomic ground state in the presence of a non-zero magnetic field. For alkali metal vapours, including commonly used isotopes such as $^{133}$Cs, $^{85}$Rb and $^{87}$Rb, Zeeman splitting is approximately linear in typical geophysical fields, and the precession frequency is related proportionally to the measured field by the gyromagnetic ratio $\gamma$. Optical detection of the spin precession allows closure of a feedback loop, in which this oscillating signal is applied to the atomic vapour via modulation of the local magnetic field \cite{Groeger2006}, the rate of optical pumping \cite{Pustelny2008}, or the polarisation of pump light \cite{Breschi2014}. Under the condition that the modulation phase matches that required for coherent feedback, the system undergoes a driven resonant response, strongly amplifying signals at the precession frequency. Such a system can be considered an alkali spin maser \cite{Chalupczak2015}.

Precessing geophysical magnetometers benefit from the atomic transduction of measured field to frequency, a physical property which can be determined precisely. Limiting noise sources on the precessing signal, such as photon shot noise, are white, meaning that the variance on this frequency (and hence magnetic field) measurement scales inversely with the cube of measurement time \cite{Jaufenthaler2021} (for derivation see \cite{Rife1974}, equations 1-17). Coupled with the high signal-to-noise measurement of optical rotation in the presence of a polarised optically pumped atomic sample, highly precise magnetometry may be achieved in geophysical magnetic fields \cite{Hunter2018}, with applications such as magneto-encephalography \cite{Limes2020} and GPS-denied navigation \cite{Canciani2021}. Furthermore, the direct mapping of precession frequency to magnetic field allows rapid changes in field to be resolved with high bandwidth, at rates even exceeding the precession frequency with use of signal phase estimation \cite{Wilson2020}. For alkali metals in geophysical magnetic fields, the magnetic precession (Larmor) frequency ranges between 70~kHz and 450~kHz. This frequency range is suitable for digitisation and real-time processing, which we exploit in this work.

Precise geophysical-field measurements are an exciting tool in several applications, especially if achieved using compact scalable hardware. Examples include detection of geomagnetic Alfv\'en resonances \cite{Beggan2018}, GPS-denied navigation \cite{Canciani2021}, and a host of terrestrial and space-based geomagnetic survey applications \cite{Gooneratne2017, Becker1995, Thebault2010}.

\textbf{Determination of magnetic field.} Consider the signal output of a spin system precessing at a magnetic resonance frequency $f_L = \gamma B_0$, where $\gamma$ is the gyromagnetic ratio, for convenience rescaled to units of Hz/nT. The lowest variance on the measurement of this frequency, in the presence of white noise, is given by the Cram\'er-Rao lower bound (CRLB) \cite{Jaufenthaler2021}
\begin{align}
\sigma^2_{\textrm{CRLB}} = \frac{3\rho^2}{\pi^2 A^2 \tau^3},
\label{CRLB}
\end{align}
where $\rho$ is the spectral noise density (units V$.$Hz$^{-1/2}$) at the measurement frequency $f_L$, $A$ the signal amplitude (V) and $\tau$ the measurement duration (s), which we assume to greatly exceed the data acquisition sample interval. The resolution of a single measurement of magnetic field is therefore given (in nT) by
\begin{align}
\Delta B_{\textrm{min}} = \frac{\sqrt{3}\rho}{\gamma \pi A \tau^{3/2}}.
\label{dB}
\end{align}
Consider the measurement of a slowly varying magnetic field $B_0$, the magnitude of which we determine through $N$ independent measurements of duration $\tau$, giving a bandwidth of $(2\tau)^{-1}$. The resolution of each measurement is given by Equation \ref{dB}, and we can find that the bandwidth-adjusted noise floor, commonly referred to as the sensitivity of the magnetometer, is proportional to $\tau^{-1}$, and given by 
\begin{align}
\delta B_{\textrm{min}} = \frac{\sqrt{6}\rho}{\gamma \pi A \tau}
\label{sensitivity}
\end{align}
in units nT.Hz$^{-1/2}$.

In the case of a freely precessing magnetometer, the atomic coherence time $T_2$ limits the duration of each measurement to $\tau \simeq 2T_2$, fixing the minimum noise floor to this physical property of the atomic system. By contrast, a continuously oscillating signal may be sampled at a rate chosen to optimally cover the desired signal bandwidth and hence achieve the highest possible sensitivity to the desired signal source. 

To achieve such a continuous atomic spin resonance, alkali spin masers have been realised using electronic feedback, using analogue phase shift and homodyne detection \cite{Chalupczak2015}. However, to reduce reliance on manually tuned analogue electronics, we use a continuous low-latency digital filter to drive resonant atomic spin precession in a $^{133}$Cs sample, implemented in firmware running on a field-programmable gate array (FPGA). In a practical sensor this represents a more scalable, flexible and portable option than analogue electronic feedback. This approach is particularly powerful in situations where the resonance phase condition is not known, for example in arbitrary geometries of the magnetic field or optical polarisation \cite{Ingleby2017}. We implement a pair of matched finite-impulse response (FIR) filters, exploiting their fixed-phase output to generate the full analytic signal, from which a frequency-agnostic constant-phase-shifted signal can be generated and applied as magnetic feedback to drive the $^{133}$Cs spins. We describe the spontaneous emergence and amplification of resonant dynamics in this digital-atomic system, measure the signal-to-noise achieved in a compact sensor package, and characterise the system bandwidth and uniformity of response to rapid variation in the measured magnetic field. 

\section{Results}\label{sec_results}

\begin{figure}[h]
\centering
\includegraphics[width=0.4\textwidth]{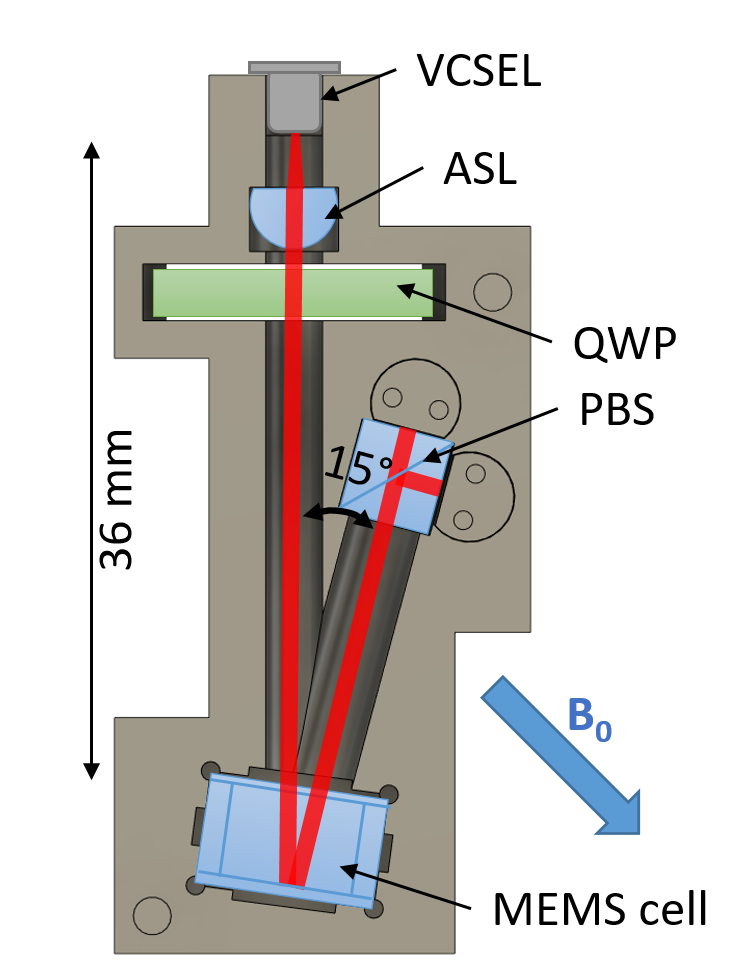}
\caption{Schematic showing the sensor package, including vertical cavity surface-emitting laser (VCSEL), aspheric lens (ASL), quarter-wave plate (QWP), polarising beam splitter (PBS) and micro-fabricated caesium vapour cell (MEMS cell). The distance between the VCSEL and the caesium cell (36~mm) and the opening angle for the reflected light (15~degrees) are indicated. The static magnetic field is offset from the incident light propagation direction by 45 degrees and the magnetic modulation field (not shown) is applied perpendicular to the plane of light propagation.}
\label{figschematic}
\end{figure}

A detailed description of the spin maser package (Figure \ref{figschematic}) and the firmware design for resonant $B_{RF}$ feedback is given in Section \ref{sec_method}.

\begin{figure}[h]
\centering
\includegraphics[width=0.8\textwidth]{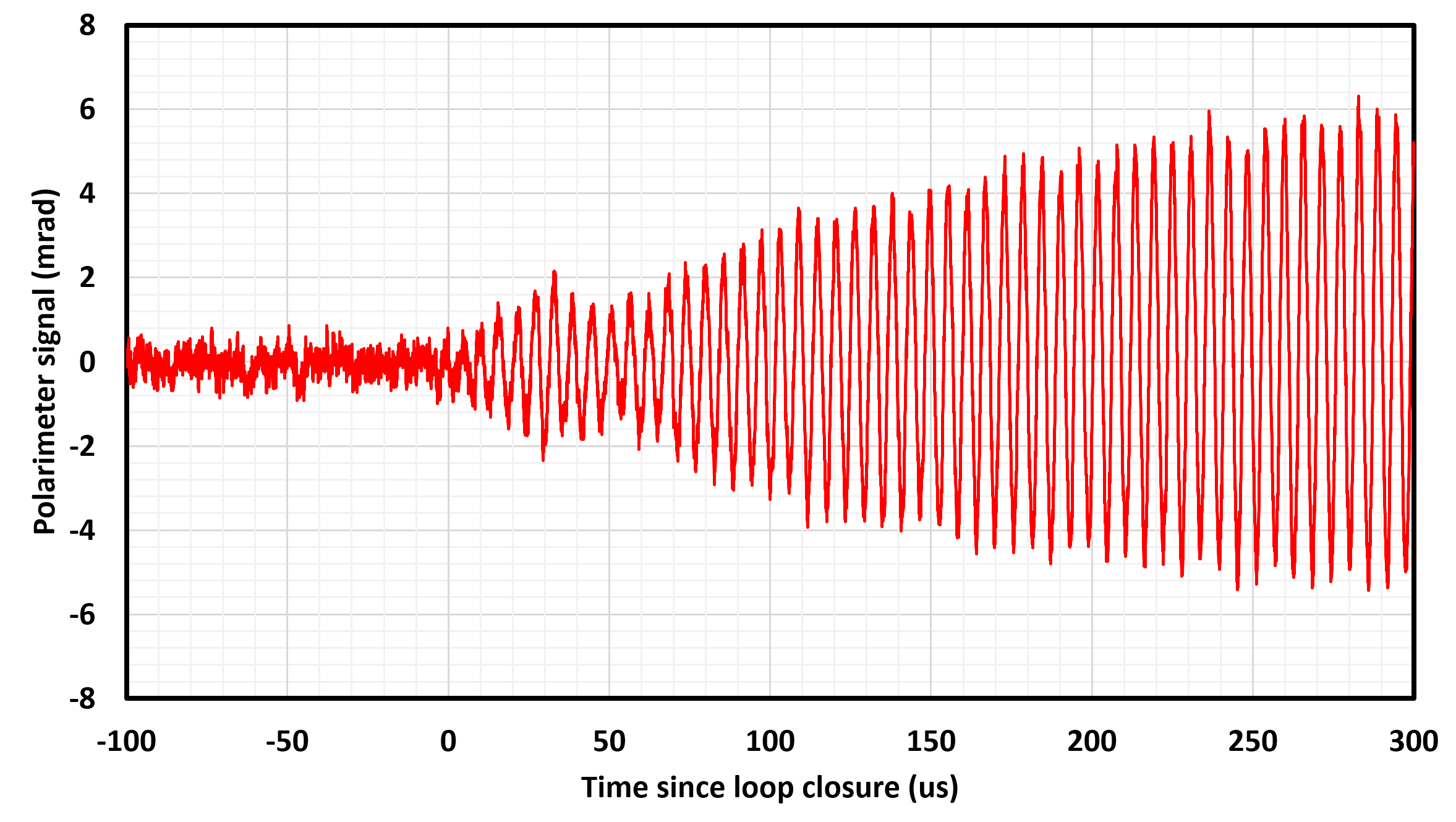}
\caption{Digital spin maser signal response following closure of feedback loop at $t=0$~s in a static $B_0$ field of 50~$\mu$T.}
\label{figswitch}
\end{figure}

\textbf{Digital spin maser resonance.} Figure \ref{figswitch} shows the polarimeter signal response following closure of the FPGA dual-FIR filter feedback loop between the polarimeter signal and the $B_{RF}$ feedback coil, while the sensor is held in a shielded magnetic field $B_0=$~50~$\mu$T. This feedback is tuned on at $t=0$~s and within 300~$\mu$s an oscillating signal is spontaneously generated at the $^{133}$Cs Larmor frequency $\sim$175~kHz.

\begin{figure}[h]
\centering
\includegraphics[width=0.8\textwidth]{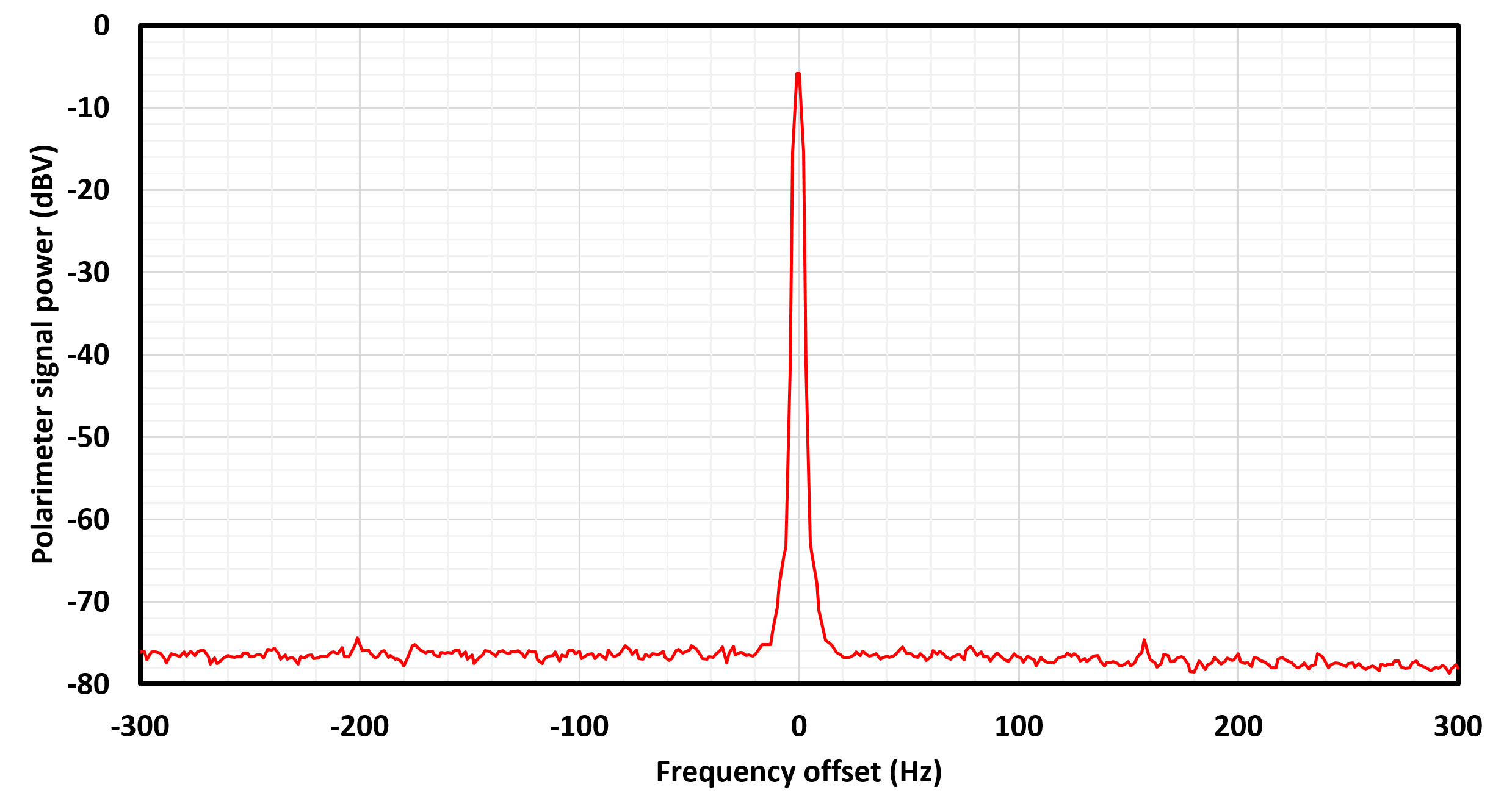}
\caption{Spectrum of digital spin maser signal around resonance in the presence of a static $B_0$ field of 50~$\mu$T, acquired using a Picoscope 5444D oscilloscope (15-bit, 200~MHz bandwidth down-sampled to 200 kHz, 2$^{17}$ samples, 32 averages, Blackman windowed). Peak power 70~dB above a white noise floor can be seen, giving a signal-to-noise-density ratio of $A/\rho=$ 3162~Hz$^\frac{1}{2}$. }
\label{figsnr}
\end{figure}

Figure \ref{figsnr} shows the spectrum around the magnetic resonance frequency for steady-state operation in a stable $B_0$ field of 50~$\mu$T, allowing the ratio of signal amplitude to noise density $A/\rho$ to be estimated. The observation of 70~dB spurious-free dynamic range (SFDR) ($A/\rho=$ 3162~Hz$^\frac{1}{2}$), allows the optimum magnetic resolution $\Delta B_\textrm{min}$ and sensitivity $\delta B_\textrm{min}$ to be estimated for a measurement time $\tau=$~1~s using Equations \ref{dB} and \ref{sensitivity} respectively. We obtain $\Delta B_\textrm{min} =$~50~fT and $\delta B_\textrm{min}=$~70~fT.Hz$^{-1/2}$ respectively, representing the CRLB-limited resolution of a 1~s duration measurement and the corresponding bandwidth-adjusted sensitivity of this measurement. 

\textbf{Bandwidth of field response.} The low-latency feedback circuit allows very rapid response between changes in the magnetic field and the observed resonance frequency. The upper limit on this response is given by the phase update rate of 1~MHz. In practice, the extraction of frequency data at this rate is hard to achieve due to limitations on data transfer rates between the FPGA and analysis software. However, the real-time firmware generation of the analytic signal (see Section \ref{sec_method}) allows the real-time phase to be found on each data sample cycle (1~MHz in this case).

\begin{figure}[h]
\centering
\includegraphics[width=0.8\textwidth]{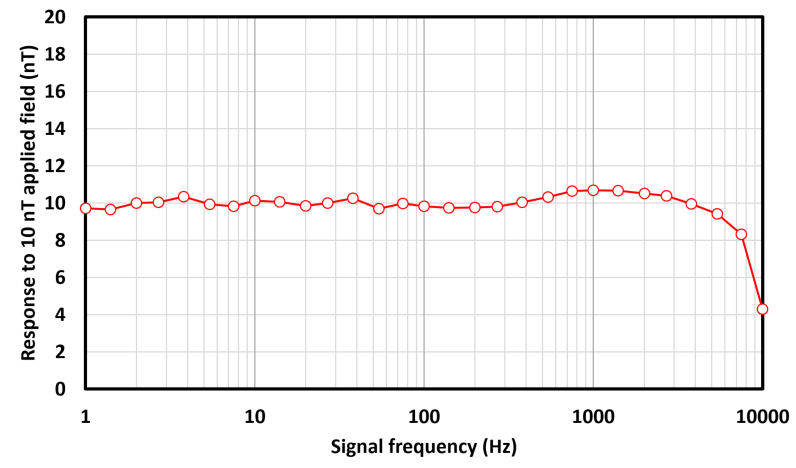}
\caption{Measured sensor signal amplitude response to 10~nT oscillating applied fields at varying frequencies up to the acquisition frequency Nyquist limit of 10~kHz (20~ks/s). The oscillating field is applied parallel to the static $B_0$ field (50~$\mu$T).}
\label{figbandwidth}
\end{figure}

From the real-time phase, the instantaneous phase step is found by numerical differentiation, and this 1~MHz data is down-sampled to 20~kHz for transfer and analysis. The phase step can then be rescaled to the instantaneous frequency, and a time series of $B_0(t)$ data recorded. A small low-impedance coil of six turns was used to generate an additional 10~nT oscillating field parallel to $B_0$, and the observed sensor response to this field extracted by demodulation of recorded $B_0(t)$ data. Figure~\ref{figbandwidth} shows this response, demonstrating uniform sensor response up to the transfer Nyquist limit of 10~kHz, with the expected amplitude reduction due to under-sampling around the Nyquist frequency visible.

\section{Discussion}\label{sec_discuss}
\textbf{Self-oscillating spin resonance.} We have confirmed that resonant self-oscillating behaviour emerges in the digital-atomic system following the activation of the feedback loop (Figure \ref{figswitch}). The response of this resonant oscillator to changes in the local magnetic field is limited to 10~kHz by the Nyquist limit on data transfer (Figure \ref{figbandwidth}), indicating that atomic system response does not impose a bandwidth limit below this frequency. 

The FIR filter architecture provides a flexible and powerful tool for defining the frequency-domain response of the system and allows a real-time estimator for the resonance frequency to be obtained. The use of a digital processor to modify and define spin maser feedback, governing and driving the inversion of Zeeman states, on a timescale faster than their precession is a valuable capability with potential for elaboration and further study. Interaction with alkali vapour ground-state Zeeman levels in geophysical fields offers a convenient sub-MHz frequency range for digital feedback and manipulation. In theses systems the use of well-established optical pumping techniques allows preparation of highly-polarised ensembles. The increasing processing power, availability and scalability of embedded processors will continuously offer enhanced impact in the manipulation of quantum systems and their application to portable devices. Development of this work has potential for study of modified spin dynamics, through infinite-impulse response (IIR) filtering, or the use of Bayesian estimation, such as Kalman filtering, to achieve CRLB-limited real-time frequency estimation. 

\textbf{Geophysical magnetic resolution.} The resolution of this digital spin maser system as a geophysical magnetometer can be estimated from the measured signal-to-noise-density (Figure \ref{figsnr}). Based on this we estimate the CRLB-limited root-mean-square resolution of this sensor to be 50~fT at 1~s sampling cadence. This measurement benefits from the persistent driven atomic oscillation, which allows $\tau$ to be chosen on the basis of signal bandwidth, rather than limited by atomic coherence $T_2$. The system also enjoys a 100\% duty cycle, without down-time for cell heating or pump laser operation. The sensor hardware used is rather modest; a single VCSEL for optical pumping and probing and a mass-produced double-pass MEMS alkali vapour cell \cite{in_prep}. This is reflected in the signal-to-noise-density ratio observed. Several well-established techniques could be used to significantly enhance signal-to-noise, such as the use of a multi-pass vapour cell \cite{Sheng2013} or the use of a re-pumping laser to achieve light-narrowing \cite{Scholtes2011}. 

\textbf{Potential applications.} The combination of these optical magnetometry techniques with the digital feedback loop demonstrated here would constitute a geophysical magnetometer of outstanding resolution, significantly exceeding parts-per-billion resolution in the Earth's field. The key advantage unlocked by the digital spin maser demonstrated here is to permit magnetic measurement by frequency estimation of a persistent resonant oscillation, not limited by atomic relaxation lifetime. In practical applications this frequency estimation may be implemented using a CRLB-limited algorithm running on a modest processor \cite{Groeger2006b}. In addition, the on-resonance phase-shift is a variable and optimised free parameter in this sensor. The capacity to vary this parameter dynamically is a unique advantage of the reported technique, offering the potential to exploit the phase-angle relations present in $B_{RF}$-modulated atomic alignment magnetometry \cite{Ingleby2017}, allowing vector readout of a single-channel geophysical optical magnetometer \cite{Ingleby2018}.

\section{Methods}\label{sec_method}
\textbf{OPM sensor head.} Since the aim of our work is the development of practical atomic sensors for real-world magnetic measurements, we have utilised a compact, fully integrated sensor module, shown in Figure \ref{figschematic}. This sensor module is built around a $^{133}$Cs atomic sample, hermetically sealed within a glass-silicon micro-fabricated cell of outer dimensions 10 x 10 x 5.5 mm \cite{in_prep}. The caesium atoms, introduced by azide decomposition, inhabit a cavity of approximate dimensions 6 x 6 x 5 mm. One wall of this cavity is formed by a silicon wafer, which is around 30\% reflective at the operating wavelength of 894.6~nm. A single-mode vertical-cavity surface emitting laser (VCSEL) of output 300~$\mu$W at this wavelength is collimated to a beam diameter of 3.2~mm, circularly polarised and used to optically pump the caesium vapour, which is heated to 80$^{\circ}$C using a non-inductively wound Ohmic heater driven at 390~kHz. The rate of depolarising caesium-wall collisions is reduced by use of 250~torr N$_2$ buffer/quenching gas, and for optimum magnetic resonance amplitude, the VCSEL is 15~GHz blue-detuned from the unresolved D1 absorption line of the caesium vapour. The oscillating optical rotation induced by the atomic sample is detected by polarisation measurement of the transmitted light using a polarimeter, comprising a polarising beam-splitter and differential photodiode circuit. Magnetic modulation and feedback is implemented using a single-turn coil to generate an oscillating field $B_{\textrm{RF}}$ on the axis perpendicular to the incident and reflected light propagation.

\textbf{Test conditions.} To establish its performance under optimal conditions, the entire sensor module is placed within a degaussed five-layer mu-metal magnetic shield, to eliminate external noise sources and allow a static magnetic field $B_0 \simeq$~50~$\mu$T to be applied. This field is established at an angle of 45$^\circ$ to the incident light propagation axis, the optimal operating geometry for an M$_x$ magnetometer \cite{Groeger2006}. Further studies under variable $B_0$ orientation will quantify its effect on signal amplitude and heading error. We note useful prior work on $B_0$ orientation effects on similar sensor configurations \cite{Lee2021}. Under optical pumping from the incident light, a net spin orientation moment is generated configurations. within the atomic sample, which precesses about $B_0$ at a frequency $f_L = \gamma B_0$. With the application of a small $B_\textrm{RF}$ field at this frequency the atomic coherence can be driven and maintained, provided a $\pi/2$ radian phase shift is maintained between the phase of $B_\textrm{RF}$ and the polarimeter signal. Under transformation to a frame co-rotating about $B_0$ with $B_\textrm{RF}$ the magnitude of the spin coherence, which is quasi-static in this frame, increases up to a maximum limited by $B_\textrm{RF}$ saturation, which dominates for $\gamma B_\textrm{RF} > T_2^{-1}$ \cite{Groeger2006}. In this frame the maximised static spin coherence corresponds to a driven inversion of the ground-state Zeeman populations away from their equilibrium unpolarised values.

\textbf{Filter design and firmware implementation.} The generation of this phase-coherent feedback in a frequency-invariant manner from the polarimeter signal is achieved using dual FIR filters to generate both a filtered signal and its Hilbert transform. This exploits the symmetries of linear-phase FIR filters, in this case, odd-coefficient even symmetry (Type 1) and odd symmetry (Type 3) filters. The phase response of a Type 3 FIR filter will be $\pi/2$ radian offset from that of a Type 1 filter of equal coefficient vector length \cite{Proakis2007}.

\begin{figure}[h]
\centering
\includegraphics[width=0.8\textwidth]{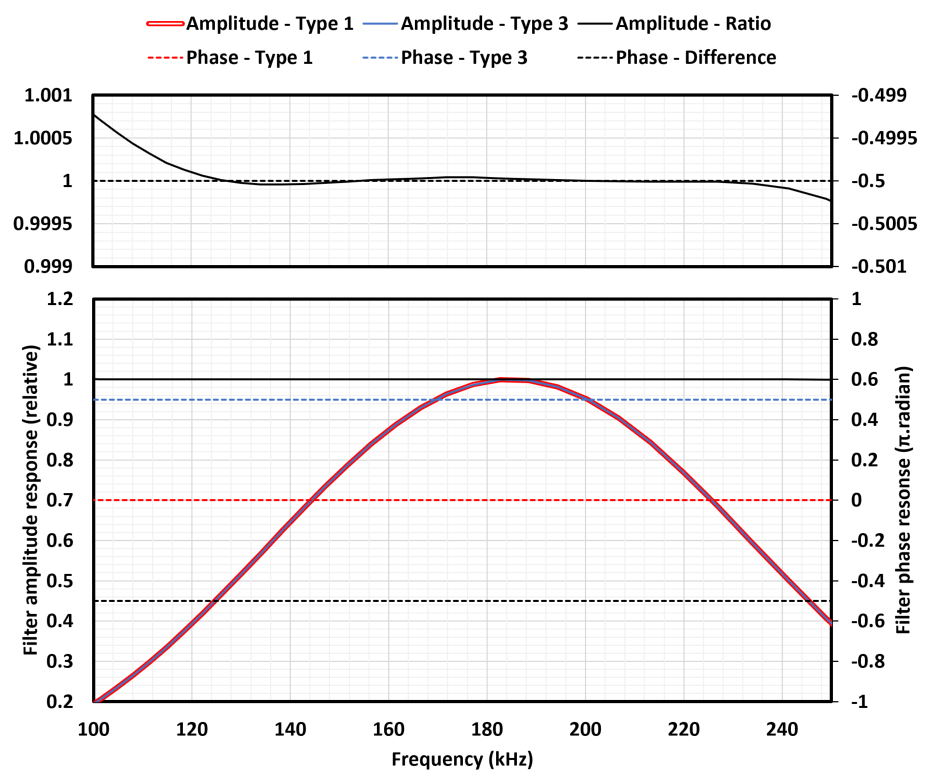}
\caption{Calculated results for the amplitude and phase frequency response of the dual FIR filters, including the effects of fixed-point signal representation. The solid curves show the relative amplitude response of the matched type 1 (red) and type 3 (blue) filters, and the ratio of their amplitudes (black). The dashed lines show the phase shift of the type 1 (red) and type 3 (blue) filters, along with their difference (black). The upper plot shows the amplitude ratio and phase difference in detail. Note that the phase difference is fixed to $\pi/2$~radian by filter coefficient symmetry.}
\label{fig_filter}
\end{figure}

The design of these dual filters was optimised in the following way. The passband was defined as 150 - 220~kHz to cover a suitable range for a $^{133}$Cs spin maser operating in geophysical fields around latitudes of 56$^\circ$~N, and the filter performance was optimised over a wider band of 0.001 - 400~kHz. The method for FIR filter design described here would work for any geophysical passband, allowing sensor operation at different latitude ranges to be achieved with firmware changes. The signal fixed-point representation was defined as 16 bits to match the analogue-digital converter (ADC) resolution, and the ADC sampling rate was specified (1~MS/s). For a given length $n$ of FIR coefficient vector, a windowed filter design tool was used to generate fixed-point coefficients for a Type 1 band-pass coefficient vector $h_1(n)$. The amplitude and phase response of filter $h_1(n)$ was found by Fourier transformation, including the effects of fixed-point representation rounding errors. The required $\pi/2$ radian phase difference of the Type 3 filter was locked by symmetry but the amplitude response was a function of the $(n-1)/2$ free parameters in the Type 3 coefficient vector $h_3(n)$. $h_3(n)$ was found iteratively by minimising the root-mean-square residual in the ratio of the amplitude response of $h_1(n)$ to $h_3(n)$ integrated over 0.001 - 400~kHz using a Levenberg–Marquardt algorithm. The coefficient vector length $n$ was allowed to vary freely between 9 and 53, and a range of band-pass filter windows were investigated. Figure \ref{fig_filter} shows the amplitude and phase response of the optimum filter, generated with $n=$~23~coefficients and an Exact Blackman window function. A root-mean-square ratio residual of 2.23~$\times$~10$^{-5}$ between the amplitude response of the Type 1 and Type 3 filters was obtained.

The output of the Type 1 and Type 3 filters can be considered the real and imaginary parts of the complex-plane analytic signal, allowing synchronous calculation of the signal phase by trigonometry on each ADC clock cycle. The live signal phase thus obtained is represented by a signed 18-bit fixed point number representing a phase between -$\pi$ and $\pi$~radian. This choice of fixed-point representation for signal phase allows phase wrapping to be unwrapped without additional operation or resolution loss by fixed point overflow. The feedback phase is found by manually shifting the signal phase by $\pi/2$~radian, and a CORDIC trigonometry algorithm used to generate phase-coherent $B_{RF}$ modulation (with a fixed amplitude below the saturation threshold), which is output to a single-turn coil at the magnetometer cell using an digital-to-analogue converter (DAC) and buffer amplifier. The live phase is also numerically differentiated to obtain a live estimate for the oscillation frequency, as described in Section \ref{sec_results}.

\backmatter


\bmhead{Acknowledgments}
The authors would like to thank Iain Chalmers for his valuable advice and discussion in firmware design. This work was funded by the UK Quantum Technology Hub in Sensing and Timing, EPSRC (Grant No. EP/T001046/1). 

\bmhead{Author contributions}
S.I. designed the system and conducted the testing, with input from P.G. and E.R. T.D. designed and fabricated the Cs vapour cell. M.M. developed and built test hardware, including bias field current supplies. All authors discussed the results and commented on the manuscript.

\bmhead{Competing interests}
The authors declare no competing interests.

\bmhead{Data availability}
The datasets used in this work are available at https://doi.org/10.15129/e6e558c9-135d-431e-a8f2-63d0fa647fbb.

\if false
\section*{Declarations}

Some journals require declarations to be submitted in a standardised format. Please check the Instructions for Authors of the journal to which you are submitting to see if you need to complete this section. If yes, your manuscript must contain the following sections under the heading `Declarations':

\begin{itemize}
\item Funding
\item Conflict of interest/Competing interests (check journal-specific guidelines for which heading to use)
\item Ethics approval 
\item Consent to participate
\item Consent for publication
\item Availability of data and materials
\item Code availability 
\item Authors' contributions
\end{itemize}

\noindent
If any of the sections are not relevant to your manuscript, please include the heading and write `Not applicable' for that section. 

\bigskip
\begin{flushleft}%
Editorial Policies for:

\bigskip\noindent
Springer journals and proceedings: \url{https://www.springer.com/gp/editorial-policies}

\bigskip\noindent
Nature Portfolio journals: \url{https://www.nature.com/nature-research/editorial-policies}

\bigskip\noindent
\textit{Scientific Reports}: \url{https://www.nature.com/srep/journal-policies/editorial-policies}

\bigskip\noindent
BMC journals: \url{https://www.biomedcentral.com/getpublished/editorial-policies}
\end{flushleft}
\fi


\bibliography{sn-bibliography}


\begin{thebibliography}{24}
\ifx \bisbn   \undefined \def \bisbn  #1{ISBN #1}\fi
\ifx \binits  \undefined \def \binits#1{#1}\fi
\ifx \bauthor  \undefined \def \bauthor#1{#1}\fi
\ifx \batitle  \undefined \def \batitle#1{#1}\fi
\ifx \bjtitle  \undefined \def \bjtitle#1{#1}\fi
\ifx \bvolume  \undefined \def \bvolume#1{\textbf{#1}}\fi
\ifx \byear  \undefined \def \byear#1{#1}\fi
\ifx \bissue  \undefined \def \bissue#1{#1}\fi
\ifx \bfpage  \undefined \def \bfpage#1{#1}\fi
\ifx \blpage  \undefined \def \blpage #1{#1}\fi
\ifx \burl  \undefined \def \burl#1{\textsf{#1}}\fi
\ifx \doiurl  \undefined \def \doiurl#1{\url{https://doi.org/#1}}\fi
\ifx \betal  \undefined \def \betal{\textit{et al.}}\fi
\ifx \binstitute  \undefined \def \binstitute#1{#1}\fi
\ifx \binstitutionaled  \undefined \def \binstitutionaled#1{#1}\fi
\ifx \bctitle  \undefined \def \bctitle#1{#1}\fi
\ifx \beditor  \undefined \def \beditor#1{#1}\fi
\ifx \bpublisher  \undefined \def \bpublisher#1{#1}\fi
\ifx \bbtitle  \undefined \def \bbtitle#1{#1}\fi
\ifx \bedition  \undefined \def \bedition#1{#1}\fi
\ifx \bseriesno  \undefined \def \bseriesno#1{#1}\fi
\ifx \blocation  \undefined \def \blocation#1{#1}\fi
\ifx \bsertitle  \undefined \def \bsertitle#1{#1}\fi
\ifx \bsnm \undefined \def \bsnm#1{#1}\fi
\ifx \bsuffix \undefined \def \bsuffix#1{#1}\fi
\ifx \bparticle \undefined \def \bparticle#1{#1}\fi
\ifx \barticle \undefined \def \barticle#1{#1}\fi
\bibcommenthead
\ifx \bconfdate \undefined \def \bconfdate #1{#1}\fi
\ifx \botherref \undefined \def \botherref #1{#1}\fi
\ifx \url \undefined \def \url#1{\textsf{#1}}\fi
\ifx \bchapter \undefined \def \bchapter#1{#1}\fi
\ifx \bbook \undefined \def \bbook#1{#1}\fi
\ifx \bcomment \undefined \def \bcomment#1{#1}\fi
\ifx \oauthor \undefined \def \oauthor#1{#1}\fi
\ifx \citeauthoryear \undefined \def \citeauthoryear#1{#1}\fi
\ifx \endbibitem  \undefined \def \endbibitem {}\fi
\ifx \bconflocation  \undefined \def \bconflocation#1{#1}\fi
\ifx \arxivurl  \undefined \def \arxivurl#1{\textsf{#1}}\fi
\csname PreBibitemsHook\endcsname

\bibitem{Bloom1962}
\begin{barticle}
\bauthor{\bsnm{Bloom}, \binits{A.L.}}:
\batitle{Principles of operation of the rubidium vapormagnetometer}.
\bjtitle{Appl. Opt.}
\bvolume{1}(\bissue{1}),
\bfpage{61}--\blpage{68}
(\byear{1962}).
\doiurl{10.1364/AO.1.000061}
\end{barticle}
\endbibitem

\bibitem{Lucivero2021}
\begin{barticle}
\bauthor{\bsnm{Lucivero}, \binits{V.G.}},
\bauthor{\bsnm{Lee}, \binits{W.}},
\bauthor{\bsnm{Dural}, \binits{N.}},
\bauthor{\bsnm{Romalis}, \binits{M.V.}}:
\batitle{Femtotesla direct magnetic gradiometer using a single multipass cell}.
\bjtitle{Phys. Rev. Applied}
\bvolume{15},
\bfpage{014004}
(\byear{2021}).
\doiurl{10.1103/PhysRevApplied.15.014004}
\end{barticle}
\endbibitem

\bibitem{Groeger2006}
\begin{barticle}
\bauthor{\bsnm{Groeger}, \binits{S.}},
\bauthor{\bsnm{Bison}, \binits{G.}},
\bauthor{\bsnm{Schenker}, \binits{J.-L.}},
\bauthor{\bsnm{Wynands}, \binits{R.}},
\bauthor{\bsnm{Weis}, \binits{A.}}:
\batitle{A high-sensitivity laser-pumped ${M}_{x}$ magnetometer}.
\bjtitle{Eur. Phys. J. D}
\bvolume{38},
\bfpage{239}--\blpage{247}
(\byear{2006}).
\doiurl{10.1140/epjd/e2006-00037-y}
\end{barticle}
\endbibitem

\bibitem{Pustelny2008}
\begin{barticle}
\bauthor{\bsnm{Pustelny}, \binits{S.}},
\bauthor{\bsnm{Wojciechowski}, \binits{A.}},
\bauthor{\bsnm{Gring}, \binits{M.}},
\bauthor{\bsnm{Kotyrba}, \binits{M.}},
\bauthor{\bsnm{Zachorowski}, \binits{J.}},
\bauthor{\bsnm{Gawlik}, \binits{W.}}:
\batitle{Magnetometry based on nonlinear magneto-optical rotation with
  amplitude-modulated light}.
\bjtitle{Journal of Applied Physics}
\bvolume{103}(\bissue{6}),
\bfpage{063108}
(\byear{2008}).
\doiurl{10.1063/1.2844494}
\end{barticle}
\endbibitem

\bibitem{Breschi2014}
\begin{barticle}
\bauthor{\bsnm{Breschi}, \binits{E.}},
\bauthor{\bsnm{Grujić}, \binits{Z.D.}},
\bauthor{\bsnm{Knowles}, \binits{P.}},
\bauthor{\bsnm{Weis}, \binits{A.}}:
\batitle{A high-sensitivity push-pull magnetometer}.
\bjtitle{Applied Physics Letters}
\bvolume{104}(\bissue{2}),
\bfpage{023501}
(\byear{2014}).
\doiurl{10.1063/1.4861458}
\end{barticle}
\endbibitem

\bibitem{Chalupczak2015}
\begin{barticle}
\bauthor{\bsnm{Chalupczak}, \binits{W.}},
\bauthor{\bsnm{Josephs-Franks}, \binits{P.}}:
\batitle{Alkali-metal spin maser}.
\bjtitle{Phys. Rev. Lett.}
\bvolume{115},
\bfpage{033004}
(\byear{2015}).
\doiurl{10.1103/PhysRevLett.115.033004}
\end{barticle}
\endbibitem

\bibitem{Jaufenthaler2021}
\begin{botherref}
\oauthor{\bsnm{Jaufenthaler}, \binits{A.}},
\oauthor{\bsnm{Kornack}, \binits{T.}},
\oauthor{\bsnm{Lebedev}, \binits{V.}},
\oauthor{\bsnm{Limes}, \binits{M.E.}},
\oauthor{\bsnm{Körber}, \binits{R.}},
\oauthor{\bsnm{Liebl}, \binits{M.}},
\oauthor{\bsnm{Baumgarten}, \binits{D.}}:
Pulsed optically pumped magnetometers: Addressing dead time and bandwidth for
  the unshielded magnetorelaxometry of magnetic nanoparticles.
Sensors
\textbf{21}(4)
(2021).
\doiurl{10.3390/s21041212}
\end{botherref}
\endbibitem

\bibitem{Rife1974}
\begin{barticle}
\bauthor{\bsnm{Rife}, \binits{D.}},
\bauthor{\bsnm{Boorstyn}, \binits{R.}}:
\batitle{Single tone parameter estimation from discrete-time observations}.
\bjtitle{IEEE Transactions on Information Theory}
\bvolume{20}(\bissue{5}),
\bfpage{591}--\blpage{598}
(\byear{1974}).
\doiurl{10.1109/TIT.1974.1055282}
\end{barticle}
\endbibitem

\bibitem{Hunter2018}
\begin{barticle}
\bauthor{\bsnm{Hunter}, \binits{D.}},
\bauthor{\bsnm{Piccolomo}, \binits{S.}},
\bauthor{\bsnm{Pritchard}, \binits{J.D.}},
\bauthor{\bsnm{Brockie}, \binits{N.L.}},
\bauthor{\bsnm{Dyer}, \binits{T.E.}},
\bauthor{\bsnm{Riis}, \binits{E.}}:
\batitle{Free-induction-decay magnetometer based on a microfabricated cs vapor
  cell}.
\bjtitle{Phys. Rev. Applied}
\bvolume{10},
\bfpage{014002}
(\byear{2018}).
\doiurl{10.1103/PhysRevApplied.10.014002}
\end{barticle}
\endbibitem

\bibitem{Limes2020}
\begin{barticle}
\bauthor{\bsnm{Limes}, \binits{M.E.}},
\bauthor{\bsnm{Foley}, \binits{E.L.}},
\bauthor{\bsnm{Kornack}, \binits{T.W.}},
\bauthor{\bsnm{Caliga}, \binits{S.}},
\bauthor{\bsnm{McBride}, \binits{S.}},
\bauthor{\bsnm{Braun}, \binits{A.}},
\bauthor{\bsnm{Lee}, \binits{W.}},
\bauthor{\bsnm{Lucivero}, \binits{V.G.}},
\bauthor{\bsnm{Romalis}, \binits{M.V.}}:
\batitle{Portable magnetometry for detection of biomagnetism in ambient
  environments}.
\bjtitle{Phys. Rev. Applied}
\bvolume{14},
\bfpage{011002}
(\byear{2020}).
\doiurl{10.1103/PhysRevApplied.14.011002}
\end{barticle}
\endbibitem

\bibitem{Canciani2021}
\begin{botherref}
\oauthor{\bsnm{Canciani}, \binits{A.J.}}:
Magnetic navigation on an {F}-16 aircraft using online calibration.
IEEE Transactions on Aerospace and Electronic Systems,
1--1
(2021).
\doiurl{10.1109/TAES.2021.3101567}
\end{botherref}
\endbibitem

\bibitem{Wilson2020}
\begin{barticle}
\bauthor{\bsnm{Wilson}, \binits{N.}},
\bauthor{\bsnm{Perrella}, \binits{C.}},
\bauthor{\bsnm{Anderson}, \binits{R.}},
\bauthor{\bsnm{Luiten}, \binits{A.}},
\bauthor{\bsnm{Light}, \binits{P.}}:
\batitle{Wide-bandwidth atomic magnetometry via instantaneous-phase retrieval}.
\bjtitle{Phys. Rev. Research}
\bvolume{2},
\bfpage{013213}
(\byear{2020}).
\doiurl{10.1103/PhysRevResearch.2.013213}
\end{barticle}
\endbibitem

\bibitem{Beggan2018}
\begin{barticle}
\bauthor{\bsnm{Beggan}, \binits{C.D.}},
\bauthor{\bsnm{Musur}, \binits{M.}}:
\batitle{Observation of ionospheric {A}lfvén resonances at 1–30 {H}z and
  their superposition with the {S}chumann resonances}.
\bjtitle{Journal of Geophysical Research: Space Physics}
\bvolume{123}(\bissue{5}),
\bfpage{4202}--\blpage{4214}
(\byear{2018})
{\href{https://arxiv.org/abs/https://agupubs.onlinelibrary.wiley.com/doi/pdf/10.1029/2018JA025264}{{https://agupubs.onlinelibrary.wiley.com/doi/pdf/10.1029/2018JA025264}}}.
\doiurl{10.1029/2018JA025264}
\end{barticle}
\endbibitem

\bibitem{Gooneratne2017}
\begin{botherref}
\oauthor{\bsnm{Gooneratne}, \binits{C.P.}},
\oauthor{\bsnm{Li}, \binits{B.}},
\oauthor{\bsnm{Moellendick}, \binits{T.E.}}:
Downhole applications of magnetic sensors.
Sensors
\textbf{17}(10)
(2017).
\doiurl{10.3390/s17102384}
\end{botherref}
\endbibitem

\bibitem{Becker1995}
\begin{barticle}
\bauthor{\bsnm{Becker}, \binits{H.}}:
\batitle{From nanotesla to picotesla — a new window for magnetic prospecting
  in archaeology}.
\bjtitle{Archaeological Prospection}
\bvolume{2}(\bissue{4}),
\bfpage{217}--\blpage{228}
(\byear{1995}).
\doiurl{10.1002/1099-0763(199512)2:4<217::AID-ARP6140020405>3.0.CO;2-U}
\end{barticle}
\endbibitem

\bibitem{Thebault2010}
\begin{barticle}
\bauthor{\bsnm{Th\'ebault}, \binits{E.}},
\bauthor{\bsnm{Purucker}, \binits{M.}},
\bauthor{\bsnm{Whaler}, \binits{K.A.}},
\bauthor{\bsnm{Langlais}, \binits{B.}},
\bauthor{\bsnm{Sabaka}, \binits{T.J.}}:
\batitle{The magnetic field of the {E}arth’s lithosphere}.
\bjtitle{Space Science Reviews}
\bvolume{155},
\bfpage{95}--\blpage{127}
(\byear{2010}).
\doiurl{10.1007/s11214-010-9667-6}
\end{barticle}
\endbibitem

\bibitem{Ingleby2017}
\begin{barticle}
\bauthor{\bsnm{Ingleby}, \binits{S.J.}},
\bauthor{\bsnm{O'Dwyer}, \binits{C.}},
\bauthor{\bsnm{Griffin}, \binits{P.F.}},
\bauthor{\bsnm{Arnold}, \binits{A.S.}},
\bauthor{\bsnm{Riis}, \binits{E.}}:
\batitle{Orientational effects on the amplitude and phase of polarimeter
  signals in double-resonance atomic magnetometry}.
\bjtitle{Phys. Rev. A}
\bvolume{96},
\bfpage{013429}
(\byear{2017}).
\doiurl{10.1103/PhysRevA.96.013429}
\end{barticle}
\endbibitem

\bibitem{in_prep}
\begin{botherref}
{T}. {D}yer et al., article in preparation.
\end{botherref}
\endbibitem

\bibitem{Sheng2013}
\begin{barticle}
\bauthor{\bsnm{Sheng}, \binits{D.}},
\bauthor{\bsnm{Li}, \binits{S.}},
\bauthor{\bsnm{Dural}, \binits{N.}},
\bauthor{\bsnm{Romalis}, \binits{M.V.}}:
\batitle{Subfemtotesla scalar atomic magnetometry using multipass cells}.
\bjtitle{Phys. Rev. Lett.}
\bvolume{110},
\bfpage{160802}
(\byear{2013}).
\doiurl{10.1103/PhysRevLett.110.160802}
\end{barticle}
\endbibitem

\bibitem{Scholtes2011}
\begin{barticle}
\bauthor{\bsnm{Scholtes}, \binits{T.}},
\bauthor{\bsnm{Schultze}, \binits{V.}},
\bauthor{\bsnm{IJsselsteijn}, \binits{R.}},
\bauthor{\bsnm{Woetzel}, \binits{S.}},
\bauthor{\bsnm{Meyer}, \binits{H.-G.}}:
\batitle{Light-narrowed optically pumped ${M}_{x}$ magnetometer with a
  miniaturized {C}s cell}.
\bjtitle{Phys. Rev. A}
\bvolume{84},
\bfpage{043416}
(\byear{2011}).
\doiurl{10.1103/PhysRevA.84.043416}
\end{barticle}
\endbibitem

\bibitem{Groeger2006b}
\begin{barticle}
\bauthor{\bsnm{Groeger}, \binits{S.}},
\bauthor{\bsnm{Bison}, \binits{G.}},
\bauthor{\bsnm{Knowles}, \binits{P.E.}},
\bauthor{\bsnm{Weis}, \binits{A.}}:
\batitle{A sound card based multi-channel frequency measurement system}.
\bjtitle{The European Physical Journal - Applied Physics}
\bvolume{33}(\bissue{3}),
\bfpage{221}--\blpage{224}
(\byear{2006}).
\doiurl{10.1051/epjap:2006020}
\end{barticle}
\endbibitem

\bibitem{Ingleby2018}
\begin{barticle}
\bauthor{\bsnm{Ingleby}, \binits{S.J.}},
\bauthor{\bsnm{O'Dwyer}, \binits{C.}},
\bauthor{\bsnm{Griffin}, \binits{P.F.}},
\bauthor{\bsnm{Arnold}, \binits{A.S.}},
\bauthor{\bsnm{Riis}, \binits{E.}}:
\batitle{Vector magnetometry exploiting phase-geometry effects in a
  double-resonance alignment magnetometer}.
\bjtitle{Phys. Rev. Applied}
\bvolume{10},
\bfpage{034035}
(\byear{2018}).
\doiurl{10.1103/PhysRevApplied.10.034035}
\end{barticle}
\endbibitem

\bibitem{Lee2021}
\begin{barticle}
\bauthor{\bsnm{Lee}, \binits{W.}},
\bauthor{\bsnm{Lucivero}, \binits{V.G.}},
\bauthor{\bsnm{Romalis}, \binits{M.V.}},
\bauthor{\bsnm{Limes}, \binits{M.E.}},
\bauthor{\bsnm{Foley}, \binits{E.L.}},
\bauthor{\bsnm{Kornack}, \binits{T.W.}}:
\batitle{Heading errors in all-optical alkali-metal-vapor magnetometers in
  geomagnetic fields}.
\bjtitle{Phys. Rev. A}
\bvolume{103},
\bfpage{063103}
(\byear{2021}).
\doiurl{10.1103/PhysRevA.103.063103}
\end{barticle}
\endbibitem

\bibitem{Proakis2007}
\begin{bbook}
\bauthor{\bsnm{Proakis}, \binits{J.G.}},
\bauthor{\bsnm{Manolakis}, \binits{D.G.}}:
\bbtitle{Digital Communications (4th Ed.)},
pp. \bfpage{669}--\blpage{766}.
\bpublisher{Pearson},
\blocation{London}
(\byear{2007}).
\bcomment{Chap. 10}.
\doiurl{10.978.1292/025735}
\end{bbook}
\endbibitem

\end{thebibliography}



\end{document}